# Mutual- and Self- Prototype Alignment for Semi-supervised Medical Image Segmentation

Zhenxi Zhang[1], Chunna Tian[1], Zhicheng Jiao[2]

[1] School of Electronic Engineering, Xidian University, Xi'an 710071, China
`zxzhang_5@stu.xidian.edu.cn`
[2] Department of Diagnostic Imaging, Brown University

**Abstract.** Semi-supervised learning methods have been explored in medical image segmentation tasks due to the scarcity of pixel-level annotation in the real scenario. Prototype alignment based consistency constraint is an intuitional and plausible solution to explore the useful information in the unlabeled data. In this paper, we propose a mutual- and self- prototype alignment (MSPA) framework to better utilize the unlabeled data. In specific, mutual-prototype alignment enhances the information interaction between labeled and unlabeled data. The mutual-prototype alignment imposes two consistency constraints in reverse directions between the unlabeled and labeled data, which enables the consistent embedding and model discriminability on unlabeled data. The proposed self-prototype alignment learns more stable region-wise features within unlabeled images, which optimizes the classification margin in semi-supervised segmentation by boosting the intra-class compactness and inter-class separation on the feature space. Extensive experimental results on three medical datasets demonstrate that with a small amount of labeled data, MSPA achieves large improvements by leveraging the unlabeled data. Our method also outperforms seven state-of-the-art semi-supervised segmentation methods on all three datasets.

**Keywords:** Prototype alignment, Semi-supervised learning, Medical image segmentation.

## 1    Introduction

Deep learning has achieved great progress in automatic medical image segmentation with many popular architectures, such as U-Net [1], V-Net [2], etc. Supervised methods rely heavily on sufficient labeled samples for the training of deep learning algorithms. Unfortunately, pixel-annotated medical images are very scarce. It motivates the development of the semi-supervised segmentation methods [3-10] by exploiting the information contained in the unlabeled samples. Self-training and consistency regularization are two main solutions in semi-supervised segmentation. Self-training based methods [3, 4] usually update and revise the pseudo-labels for several times iteratively, which is time-consuming. Recently, many efforts [6-10] have been devoted to the design of consistency regularization by enforcing the consistent predictions of different perturbations.

The mean-teacher framework [9] usually serves as the base flow to introduce consistency constraints. The student network learns from the temporally ensembled teacher network by penalizing the inconsistent predictions. To ensure the more accurate prediction from the teacher model, Yu et al. [6] estimate the uncertainty of the teacher predictions with the Monte Carlo Dropout to filter out the unreliable predictions. The feature uncertainty is also estimated in [7] to build a learnable double-uncertainty consistency loss. Although promising results have been achieved, these series of methods occupy more GPUs and have high computational costs due to multiple forward passes in the uncertainty estimation process. Luo, et al. [10] propose an uncertainty rectified pyramid consistency method, which encourages the predictions at multiple scales of one network to be consistent for the given input, which is computationally efficient with only a single pass. The multi-scale predictions have different spatial frequencies, which hampers the consistency learning and needs a specific strategy to conquer this problem. On the other hand, many consistency training methods face the trivial solution between different predictions. In this paper, our method also explores the idea of consistent training and adopt the idea of prototypical feature learning [22, 23] to conquer these problems. Besides, the proposed method differs [23] in three aspects. a) We design a more reliable prediction framework based on multiple prototype pairs, which are more representative than single prototype pair in the labeled prototype alignment process. Single prototype is prone to overfit the labeled data with limited representation. Differently, multiple prototypes are utilized to generate multiple predictions, which alleviates the overfitting and predicts more reliable results of variable unlabeled images. b) CPC uses the plain pseudo label for UPA. Differently, our label generation method uses the voting strategy based on multiple predictions, which has potential to filter out the noisy components of pseudo labels for better UPA. c) We propose a novel self-prototype alignment (SPA) method based on the voting results. This method boosts the compactness and separateness of the prototypes. CPC only has single prediction based on one prototype pair, so it is impossible for CPC to equip our SPA module.

Recently, prototypical feature learning has been widely used in different segmentation tasks, such as few-shot segmentation [11, 12], cross-domain segmentation [13, 14]. In few-shot learning, most methods usually generate one or few prototypical vectors of the object features in support images. Then, the predictions of the query images are produced by measuring the distance between the category-wise prototypes and pixel-wise features.

Inspired by this idea, each prototypical pair of all semantic classes in one image can be considered as one additional classifier. Thus, we propose a novel mutual- and self-prototype alignment framework for semi-supervised medical image segmentation, which includes a mutual-prototype alignment (MPA) cross images and a self-prototype alignment (SPA) within the image. The overview of the proposed framework is shown in Fig. 1. MPA contains the labeled prototype alignment (LPA) and unlabeled prototype alignment (UPA). The prototype-based predictions are generated by comparing the pixel-wise feature embedding to the class-specific prototypical pairs from labeled to unlabeled and from unlabeled to labeled in a mutual manner. In addition, the random interaction between unlabeled and labeled in MPA avoids the trivial solution in consistency training. With the prototype-based predictions, we propose two consistent constraints in MPA in a single forward pass. The constraints enable the model to generate more consistent prototypes of both the labeled and unlabeled. Moreover, the multiple predictions of the unlabeled image are generated by referring to the multiple labeled prototypical pairs. With the guidance of the multiple predictions, we propose SPA to promote inter-class separation and intra-class compactness by regional prototype alignment within the image. We conduct extensive experiments on three medical image datasets: ISIC 2018 dataset for skin lesion segmentation, Kvasir-SEG dataset for polyp segmentation, and RIM-ONE dataset for optic disc segmentation. The experiment results indicate that our semi-supervised method achieves large improvements by mutual-prototype alignment and self-prototype alignment, which outperforms seven state-of-the-art semi-supervised segmentation methods.

## 2 Methodology

### 2.1 Problem formulation

Under the semi-supervised setting, the training material usually includes a small set of labeled samples $\mathcal{D}_l = \{(X_{li}, Y_{li}) | i \in \{1,2,...,N_l\}\}$, where $(X_{li}, Y_{li})$ denotes the image with corresponding one-hot ground truth, and a large set of unlabeled samples $\mathcal{D}_u = \{X_{uj} | j \in 1,2,...,N_u\}$. We aim to train a more robust segmentation model with the joint exploration of the labeled and unlabeled samples. To efficiently exploit the unlabeled information, we introduce three consistency regularization losses in MSPA as depicted in Fig. 1 based on MPA and SPA. In specific, MPA includes LPA and UPA.

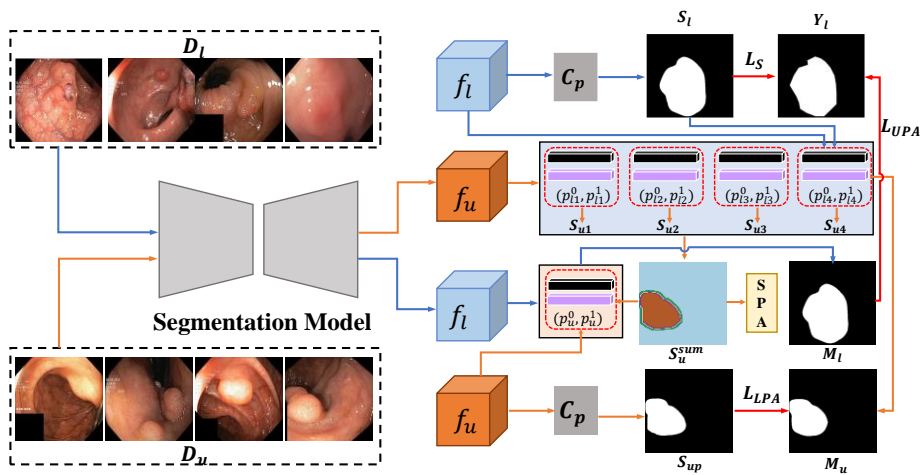

**Fig. 1.** The flowchart of the proposed MSPA framework. The blue and orange line represent the processing flow of the labeled part and unlabeled part, respectively.

## 2.2 Mutual-Prototype Alignment

**Labeled Prototype Alignment.** As shown in Fig. 1, we obtain the feature representation $f_{li} \in \mathbb{R}^{D \times H \times W}$ of the labeled image, where $D$, $H$, and $W$ represent the number of feature channels, height and width of the feature map, respectively. Given the predicted mask $S_{li}$, we generate the prototype feature $p_{li}^c \in \mathbb{R}^D$ of the labeled image $X_{li}$ for class $c$ as follows:

$$p_{li}^c = \frac{\sum_{(h,w)} f_{li}(h,w) \cdot \mathbb{1}[S_{li}(h,w) = c]}{\sum_{(h,w)} \mathbb{1}[S_{li}(h,w) = c]}, \quad (1)$$

where $(h, w)$ is the position index, $c$ represents the class which equals 0 or 1, and $\mathbb{1}[\cdot]$ is the indicator function. In one iteration, we calculate $N$ prototype pairs $\{p_{li}^0, p_{li}^1\}$ from all labeled images using Eq. 1. Each prototype pair containing global semantic information across feature channels can be seen as a classifier. Given the feature maps $f_u$ of the unlabeled image, the prototype pairs are used to locate the similar features in $f_u$ by feature matching. The cosine similarity $G_{ui}^c$ between each feature element in $f_u$ and each class prototype is calculated as follows:

$$G_{ui}^c(h,w) = \frac{f_u(h,w) \cdot p_{li}^c}{\|f_u(h,w)\| \cdot \|p_{li}^c\|}. \quad (2)$$

The multiple similarity scores for each semantic class are inferred by comparing to different prototype pairs. Then, a pixel-wise class probability of the unlabeled image is obtained by taking all similarity scores into consideration as:

$$\widehat{M}_u^c(h,w) = \frac{\exp(\sum_{i=1}^N G_{ui}^c(h,w))}{\sum_c \exp(\sum_{i=1}^N G_{ui}^c(h,w))}. \quad (3)$$

We introduce the consistency constraint $L_{\text{LPA}}$ of unlabeled predictions by minimizing the $L_2$ distance between the plaint probability prediction $\hat{S}_{up}^c$ and $\widehat{M}_u^c$ as follows:

$$L_{LPA} = \frac{1}{C \times H \times W} \sum_{h,w} \sum_c \left\| \hat{S}_{up}^c(h,w) - \widehat{M}_u^c(h,w) \right\|_2, \quad (4)$$

where $C$ denotes the number of semantic classes. Since the introduction of another prediction $\widehat{M}_u^c$ is based on prototypical feature learning from the labeled images, this module is defined as labeled prototype alignment.

**Unlabeled prototype alignment.** This module introduces the prototype alignment from unlabeled to labeled image. If the segmentation results of the unlabeled images are good enough, the prototypes extracted from the unlabeled images should segment the labeled images well. This process is defined as unlabeled prototype alignment (UPA), which boosts the semi-supervised learning in a reverse direction. Taking the similarity maps $G_{ui}^c$ as input, we obtain the binary predictions $\hat{S}_{ui} \in [0,1]^{H \times W}$ as follows:

$$S_{ui}(h,w) = \underset{c}{\mathrm{argmax}}\, G_{ui}^c(h,w), i \in \{1,2,\ldots,N\}. \tag{5}$$

Similarly, we generate the binary segmentation map $S_{up}$ based on the plaint SoftMax probability $\hat{S}_{up}^c$. For unified expression, we mark $S_{up}$ as $S_{u(N+1)}$. To enhance the reliability of the pseudo label $S_u$, we introduce a voting strategy on multiple predictions. We add up all binary predictions to get a score $S_u^{sum} \in \{0,1,2,\ldots,N+1\}$, which is formulated as:

$$S_u^{sum} = \sum_{i=1}^{N+1} S_{ui}. \tag{6}$$

In our experiment, $N$ is set to 4. Then, we adopt the voting strategy to facilitate the model to generate more accurate pseudo label $\hat{S}_u$ as:

$$S_u = \begin{cases} 1, S_u^{sum} = 3,4,5 \\ 0, S_u^{sum} = 0,1,2 \end{cases}. \tag{7}$$

Given the feature representation $f_u$ and $\hat{S}_u$ of the unlabeled image, we calculate the prototypes of the unlabeled image following Eq. 1. We also obtain $N$ prototypical pairs of different unlabeled images $\{p_{uj}^0, p_{uj}^1\}$ in one iteration, where $j \in \{1,2,\ldots,N\}$ denotes the index of unlabeled image. Then, the unlabeled prototypical pair $\{p_u^0, p_u^1\}$ is averaged by:

$$p_u^c = \frac{1}{N}\sum_{j=1}^{N} p_{uj}^c \tag{8}$$

Then, we compute the cosine similarity $G_l^c$ between each unlabeled prototype $p_u^c$ and each feature element of the labeled feature map $f_l$ following Eq. 2. Further, we apply the SoftMax function based on $G_l^c$ to produce a probability map $\widehat{M}_l^c$ over semantic classes following Eq. 3. Finally, the cross-entropy loss ($L_{ce}$) is applied for UPA as:

$$L_{UPA} = L_{ce}(\widehat{M}_l, Y_l). \tag{9}$$

## 2.3 Self-Prototype Alignment

In this subsection, we propose a self-protype alignment method within unlabeled image (see Fig. 2). Taking the score map $S_u^{sum}$ into consideration, the region whose $S_u^{sum}$ equals 3, denoted as $R_3$, contains relatively uncertain foreground predictions. The region $R_4$ is composed of relatively reliable foreground predictions. To enhance the intra-class compactness, we push the feature distribution of $R_3$ close to the feature distribution of $R_4$ by regional prototype alignment. The regional prototype is inferred in Eq. 10. $\hat{S}_{up}^c$ is the predicted class probability, which plays a soft weighting role in the regional prototype aggregation process.

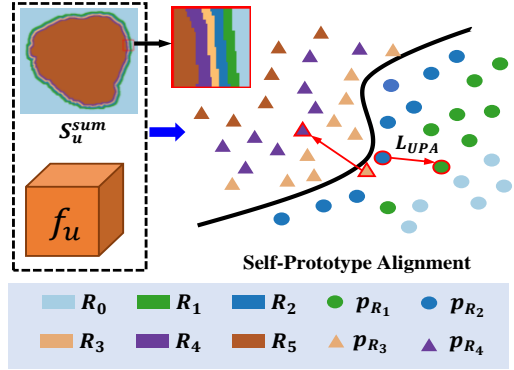

**Fig. 2.** The diagrammatic sketch of the SPA method. The black curve represents the decision boundary. SPA aims to reduce the distances between class-specific regional prototypes to regularize the feature space of the unlabeled images.

$$p_{R_k} = \frac{\sum_{(h,w)} \hat{S}_{up}^c(h,w) \cdot f_u(h,w) \cdot \mathbb{1}[S_u^{sum}(h,w) = k]}{\sum_{(h,w)} \hat{S}_{up}^c(h,w) \cdot \mathbb{1}[S_u^{sum}(h,w) = k]} \quad (10)$$

Then, the regional prototype alignment of foreground class is employed by minimizing the $L_1$ distance between $p_{R_3}$ and $p_{R_4}$. Similarly, we implement SPA of background class between $R_1$ and $R_2$. The loss function of SPA is defined as:

$$L_{RPA} = \|p_{R_3} - p_{R_4}\|_1 + \|p_{R_2} - p_{R_1}\|_1. \quad (11)$$

**Total Loss Function.** The overall loss function $L_{total}$ is the combination of the supervised loss $L_S$ and the prototype alignment loss $L_{PA}$.

$$L_{total} = L_S + \lambda \cdot L_{PA}. \quad (12)$$

In Eq. 12, $L_S = L_{ce}(\hat{S}_l, Y_l)$ represents cross-entropy loss among the probability prediction and ground truth of the labeled images, $L_{PA} = L_{LPA} + L_{UPA} + L_{SPA}$ denotes the sum of three kinds of prototype alignment losses. $\lambda(t) = w_{max} \cdot e^{(-5(1-t/t_{max})^2)}$ is a weighting function with Gaussian ramp-up curve [10] to balance the supervised learning loss and the prototype alignment loss, where $w_{max}$ is the final weight, $t$ is the current training step and $t_{max}$ is the maximum training step.

## 3 Experiments and Results

### 3.1 Dataset and Implementations

**Polyp Segmentation.** Kvasir-SEG dataset [18] contains 1000 polyp images from colonoscopy video sequences. These images are divided into 600 for training, 200 for validation and 200 for testing. All images are resized to $416 \times 416$ pixels.
**Optic Disc Segmentation.** We utilize the RIM-ONE dataset [19] to perform the optic disc segmentation task, which contains 485 images in total. We randomly select 380 images as the training set, 50 images as the validation set, and the rest for testing. All images are resized to 416 × 416 pixels.

**Skin lesion segmentation.** We use the ISIC 2018 dataset [17] to evaluate the semi-supervised segmentation performance of our method. This dataset comprises 2594 images, among which 2000 are used for training, 194 for validation, and 400 for testing. We resize all images to 384 × 512 pixels by bilinear interpolation.

We use the commonly-used Dice similarity coefficient ($Dsc$), Intersection over union ($IoU$), Sensitivity ($Sen$), Specificity ($Spe$), and pixel-level Accuracy ($Acc$) to evaluate the effectiveness of the proposed semi-supervised segmentation method.

**Implementation details.** The framework is implemented in PyTorch with a single Nvidia GeForce RTX 3090 GPU. We adopt Adam optimizer in the training procedure. The learning rate is set to 0.0001. For each iteration, four labeled samples and four unlabeled samples are fed to the segmentation model simultaneously. We apply data augmentation strategies on all datasets, including random image rescaling, random flipping, and random color distortion. We use the 2D DenseUNet architecture [20] as the backbone in all experimental groups.

**Table 1.** Ablative results of MSPA method on Kavsir-SEG dataset (%)

| Loss Function | $Dsc$ | $IoU$ | $Spe$ | $Sen$ | $Acc$ |
|---|---|---|---|---|---|
| $L_s$ | 82.88 | 75.03 | 98.06 | 79.19 | 91.39 |
| $L_s + L_{LPA}$ | 85.16 | 76.65 | 98.12 | 83.40 | 92.17 |
| $L_s + L_{LPA} + L_{UPA}$ | 86.13 | 77.89 | 98.45 | 84.13 | 92.60 |
| $L_s + L_{LPA} + L_{UPA} + L_{RPA}$ | **87.09** | **79.77** | **98.64** | **84.87** | **93.03** |

### 3.2 Ablation Study

We conduct the ablation study to verify the effectiveness of the proposed key components (LPA, UPA, and SPA) on Kavsir-SEG dataset. We employ four experimental settings by adding each module gradually. The detailed results are given in Table 1. Compared with the supervised training only learned from labeled samples, the introduction of LPA improves $Dsc$ by 2.28%, $IoU$ by 1.62%. The results suggest the consistency constraint on unlabeled data by the labeled prototypical feature learning boosts the semi-supervised segmentation performance. We further add the UPA in the third group. We find that flowing the information from unlabeled to labeled gain the improvements in terms of all metrics, where the mean $Dsc$ and $IoU$ are improved by 3.25% and 2.86% compared to the supervised counterpart in the first row of Table 1, respectively. These results indicate MPA can align the unlabeled prototypes and labeled prototypes to a consistent embedding, leading to better segmentation results. In addition, we employ SPA in the last group, constituting the overall MSPA framework. By integrating SPA, the final MSPA gains $Dsc$ improvement of 4.21% over the supervised counterpart, which indicates that the self-prototype alignment enhances the intra-class consistency by reducing the disagreements of the class-specific regional prototypes.

**Table 2.** A comparison with the state-of-the-art on Kavsir-SEG dataset (%). Note that "L" means the labeled image, and "U" means the unlabeled image.

| Method | Data | $Dsc$ | $IoU$ | $Spe$ | $Sen$ | $Acc$ |
|---|---|---|---|---|---|---|
| Supervised | 600L | 90.24 | 83.43 | 98.53 | 90.74 | 95.21 |
| Supervised | 120L | 82.88 | 75.03 | 98.06 | 79.19 | 91.39 |
| MT [9] | 120L+480U | 85.13 | 75.64 | 98.38 | 82.08 | 92.04 |
| UAMT [6] | 120L+480U | 85.21 | 77.49 | 98.30 | 81.76 | 92.26 |
| TCSM_V2 [8] | 120L+480U | 85.59 | 77.04 | 98.20 | 82.03 | 92.20 |
| DAN [16] | 120L+480U | 83.43 | 76.44 | 97.12 | 80.02 | 91.69 |
| PseudoSeg [21] | 120L+480U | 83.48 | 76.53 | 98.02 | 81.23 | 92.03 |
| E-min [15] | 120L+480U | 83.59 | 75.17 | 97.43 | 82.88 | 91.48 |
| UCPC [10] | 120L+480U | 84.10 | 76.21 | 98.12 | 83.12 | 91.67 |
| **Ours** | 120L+480U | **87.09** | **79.77** | **98.64** | **84.87** | **93.03** |

### 3.3 Comparison with Other Semi-supervised Methods

We compare the proposed framework with other existing semi-supervised segmentation methods, including MT [9], UAMT [6], TCSM_V2 [8], DAN[16], PseudoSeg [21], E-min [15], and UCPC [10]. Note that all groups use the same data augmentation methods, training strategies, and backbone model for fair comparisons. Table 2 reports the quantitative results on Kvasir-SEG dataset. Note that the comparative results on the other two datasets are reported in the supplementary material due to the space limitation. The proposed MSPA outperforms other methods in terms of all evaluation metrics consistently. In addition, our method achieves the largest improvements by 4.21%, 4.74%, and 1.64% on $Dsc$, $IoU$, and $Acc$, respectively, compared with the results achieved by the supervised training using only 120 labeled samples. It demonstrates that the mutual-prototypical feature learning and the self-prototypical feature learning method results in the performance gain, which are capable of exploring the unlabeled information effectively. We also visualize some typical segmentation results in Fig. 3 in the order of Kvasir-SEG, RIM-ONE, and ISIC 2018 dataset. It can be seen that our method locates the tissue boundaries well. Especially in the first two colonoscopy images, the polyps are hard to segment duo to the camouflage effects. Our method still achieves relatively accurate segmentation results compared to other methods.

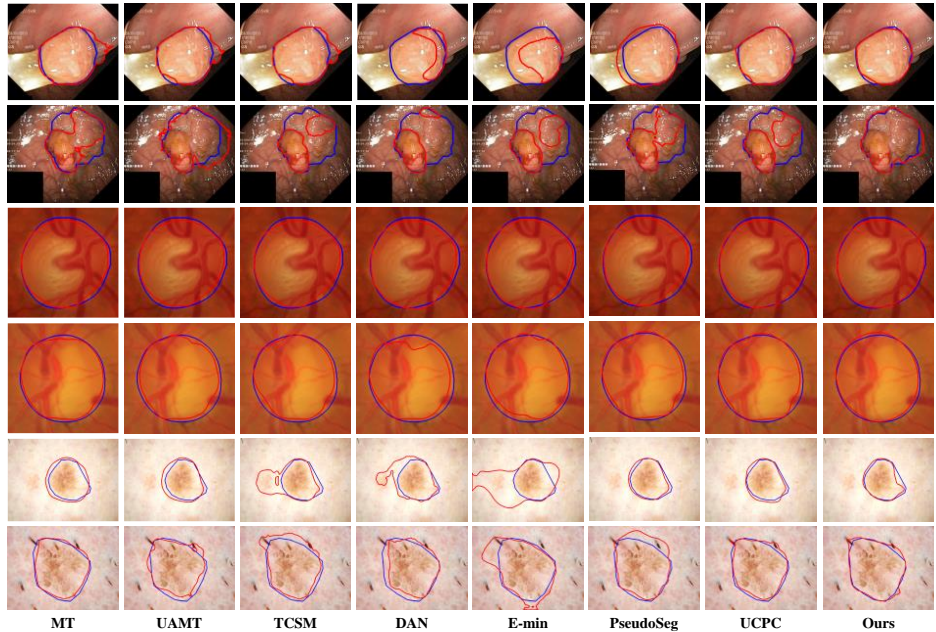

**Fig. 3.** Visualization of the typical segmentation results. Blue and red colors denote the ground truths and predictions, respectively.

## 4 Conclusion

In this paper, we study the prototypical correction learning in semi-supervised medical image segmentation. We design additional consistency constraints in semi-supervised training by introducing prototypical correction learning. A novel mutual- and self- prototype framework is proposed to exploit the unlabeled information. The mutual-prototypical feature learning effectively aligns the labeled prototypes and the unlabeled prototypes to assist the semi-supervised training. The self-prototypical feature learning is proposed to enhance the intra-class compactness on feature space. Extensive results on three datasets demonstrate that our MSPA achieves superior performance on different semi-supervised segmentation tasks. Our future work will focus on the design of the regional prototypical learning from the labeled part to the unlabeled part.